\documentclass[aps,floatfix,prd,twocolumn,a4paper,10pt,citeautoscript,
preprintnumbers,superscriptaddress,showpacs,longbibliography]{revtex4-1} 

\usepackage{graphicx,color,epsfig,rotate}
\usepackage{color}
\usepackage{graphicx}
\usepackage{dcolumn}
\usepackage{amssymb}
\usepackage{amsmath}
\usepackage{amsfonts}
\usepackage{docs}%
\usepackage{bm}%
\usepackage{wrapfig}
\usepackage[colorlinks=true,linkcolor=blue,pagecolor=blue,filecolor=blue,menucolor=blue,urlcolor=blue,citecolor=blue,anchorcolor=blue]{hyperref}%
\usepackage{graphics}
\usepackage{standalone}
\usepackage[toc,page]{appendix}
\usepackage{import}
\usepackage{enumerate}
\usepackage{sidecap}

\usepackage[english]{babel} 	
\usepackage{amsmath}        	
\usepackage{amssymb}        	
\usepackage{amsfonts}
\usepackage{bm}					
\usepackage{bbm}            	
\usepackage{empheq}
\usepackage[normalem]{ulem}

\begin{document}

\title{Flat band in disorder driven non-Hermitian Weyl semimetals}

 \author{A.~A.~Zyuzin}
  \affiliation{Department of Applied Physics, Aalto University, P.~O.~Box 15100, FI-00076 AALTO, Finland}
 \affiliation{Department of Physics, Royal Institute of Technology, SE-106 91 Stockholm, Sweden}
\affiliation{Ioffe Physical--Technical Institute,~194021 St.~Petersburg, Russia}

 \author{A.~Yu.~Zyuzin}
\affiliation{Ioffe Physical--Technical Institute,~194021 St.~Petersburg, Russia}


\begin{abstract}
We study the interplay of disorder and bandstructure topology in a Weyl semimetal with a tilted conical spectrum around the Weyl points. 
The spectrum of particles is given by the eigenvalues of a non-Hermitian matrix, which contains contributions from a Weyl Hamiltonian and complex self-energy due to electron elastic scattering on disorder. 
We find that the tilt-induced matrix structure of the self-energy gives rise to either a flat band or a nodal line segment at the interface of the electron and hole pockets in the bulk
bandstructure of type-II Weyl semimetals depending on the Weyl cone inclination.
For the tilt in a single direction in momentum space, each Weyl point expands into a flat band lying on the plane, which is transverse to the direction of the tilt. The spectrum of the flat band is fully imaginary and is separated from the in-plane dispersive part of the spectrum by the ``exceptional nodal ring" where the matrix of the Green function in momentum-frequency space is defective. The tilt in two directions might shrink a flat band into a nodal line segment with ``exceptional edge points". We discuss the connection to the non-Hermitian topological theory.
\end{abstract}
\maketitle

There are different topologically nontrivial realizations of the electronic band structure in three-dimensional systems. 
Among them are Weyl and Dirac semimetals. The band structures of these materials host electron and hole band-touching points in momentum space with a conical spectrum around them. The low-energy excitations in these systems are condensed-matter analogs of relativistic fermions, which are described by Weyl or Dirac Hamiltonians \cite{Abrikosov, Murakami, Savrasov, Burkov_Balents}. Another example is a semimetal with a line of nodes in its bandstructure \cite{Heikila_Volovik, BHB}. It was also realized that the tilt of the conical spectrum around the Weyl points in momentum space results 
in a Lifshitz transition, whereby the point like Fermi surface
transforms into electron and hole Fermi pockets connected to each other by a single point \cite{FirstWeyl2, Soluyanov}. This
is the so-called type-II phase of a Weyl semimetal. 
The classification of three-dimensional Weyl fermions was extended to interacting systems in \cite{Volovik_type3}. It was shown that the band structure of an interacting Weyl semimetal might host Weyl cones, which are surfaces separating the propagating fermionic states from the fermionic states with complex spectra. 
For a review on Weyl-Dirac semimetals see Refs. \cite{Weyl_review_burkov, Weyl_review}.

Recent research has focused attention on the topological properties of one- and two-dimensional optic \cite{NonHerm3, NonHerm4, flach} and electronic \cite{NonHerm1,NonHerm2, Shen_Fu, Kozii_Fu} band structures of the so-called non-Hermitian systems. Typical examples are the electromagnetic field in amplifying or lossy materials or interacting electronic Dirac systems.
In particular, the existence of topological ``exceptional points" and bulk Fermi arcs in band structures of Dirac materials were predicted \cite{NonHerm3, NonHerm4, flach, NonHerm1,NonHerm2, Shen_Fu, Kozii_Fu}.

The study of non-Hermitian topological objects in momentum space was extended to the three-dimensional Weyl system based on the model of a dissipative cold atomic gas \cite{Weyl_ring}. 
It is well known that the Weyl points in Hermitian systems are topologically stable in the sense, 
that any perturbation can only shift them in energy and momentum space (for a review, see Refs. \cite{Weyl_review_burkov, Weyl_review}).
Mathematically, in order to find accidental degeneracy of two eigenvalues of a $2\times 2$ Hermitian matrix
three independent real parameter must be tuned, which are three components of momentum in three-dimensional crystals. Hence Weyl points of opposite chirality can be only annihilated pairwise. 
In the non-Hermitian systems Weyl points might be unstable, since less number of parameters are required to find a level degeneracy \cite{Berry}.
Namely, it was shown that in the presence of a non-Hermitian term in the Weyl Hamiltonian, the Weyl point might transform into a flat band with an exceptional nodal ring at the edge of the band \cite{Weyl_ring}. 
Here we demonstrate that such unusual topological electronic bandstructure can be realized in disordered type-II Weyl semimetal.

\begin{figure}[t]
\begin{tabular}{cc}
\includegraphics[width=6.0cm]{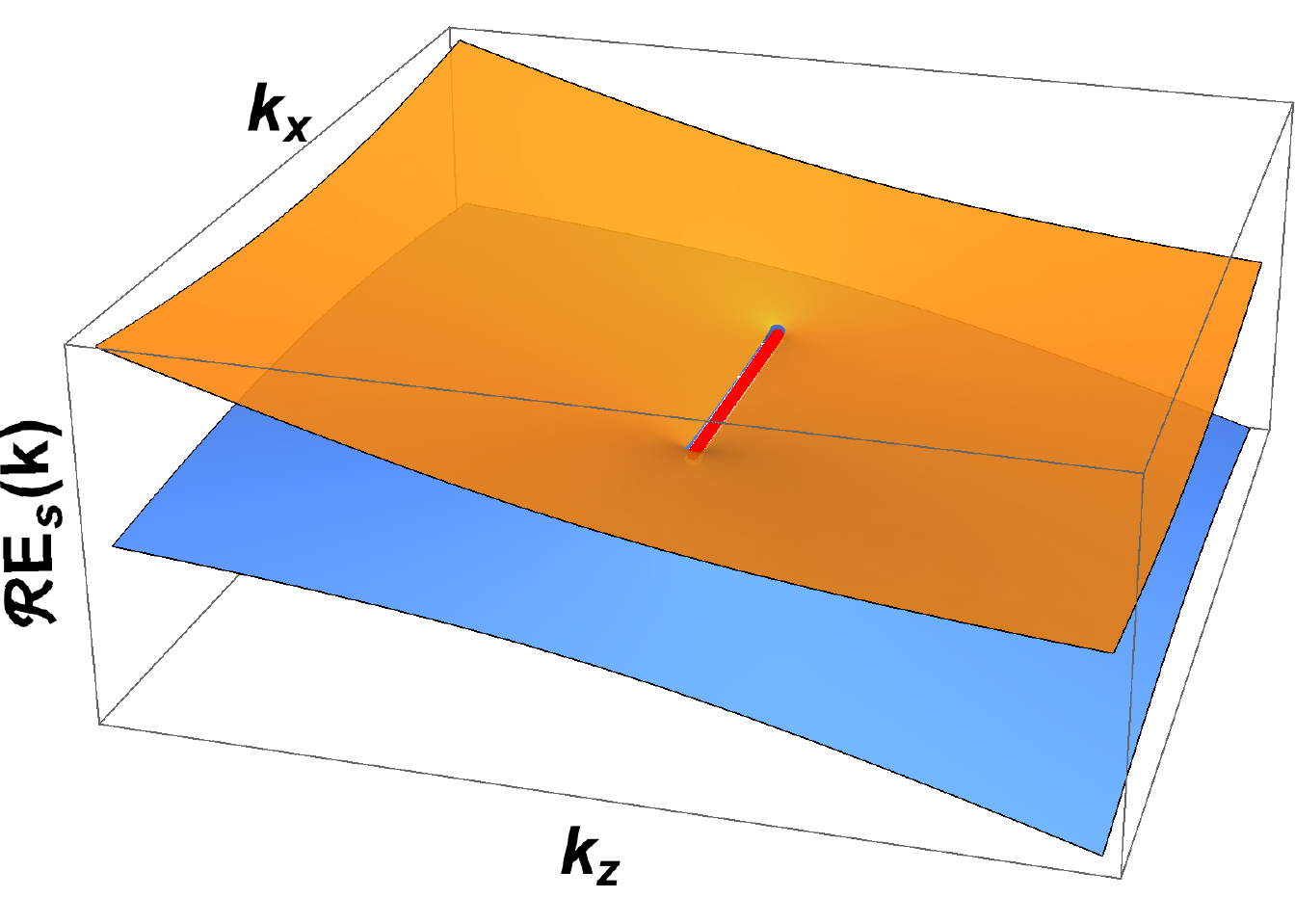}
\end{tabular}
\caption{(Color online) 
The real part of the spectrum $E_s(k_x, 0, k_z)$ of a disordered type-II Weyl semimetal. The tilting is along $\boldsymbol{e}_z$ in momentum space and $s=-1$, see Eq. \ref{Main_Result} in the text for details. 
The electron and hole pockets touch along the line segment, at which the spectrum is purely imaginary and single valued at the edge points. 
The line segment is a diameter of the circle flat band, which lies on the plane $k_z=0$ as shown in the right panel of Fig. \ref{fig2}.}
\label{fig1} 
\end{figure}

\begin{figure}[t]
\begin{tabular}{cc}
\includegraphics[width=4.5cm]{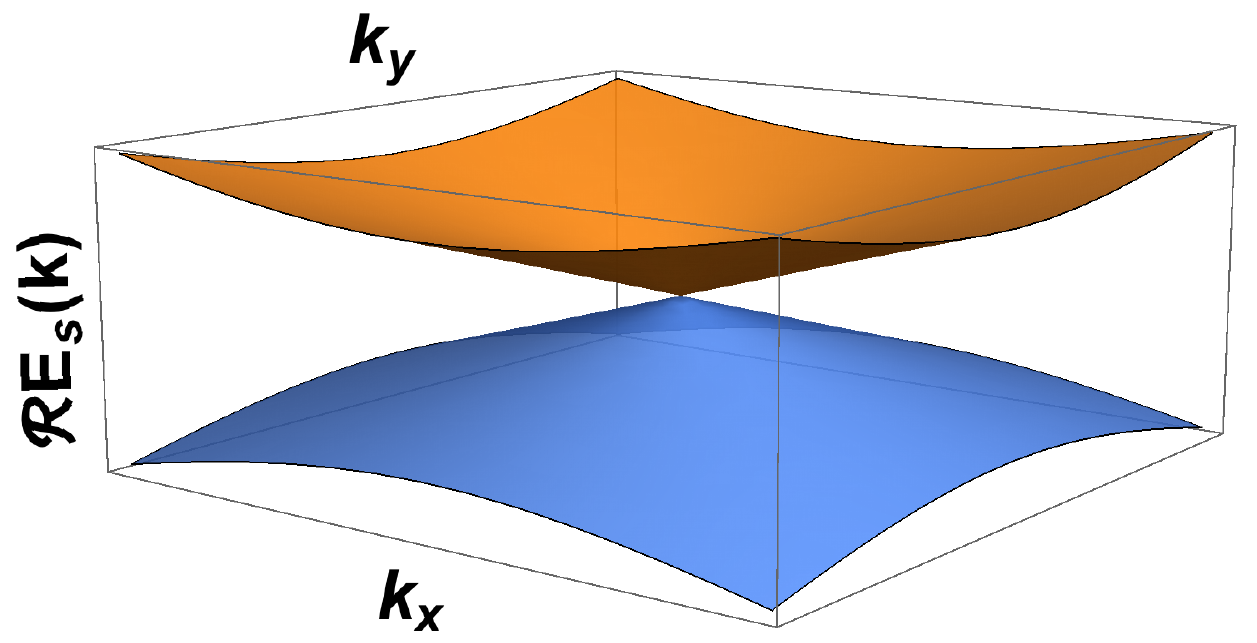} \includegraphics[width=4.5cm]{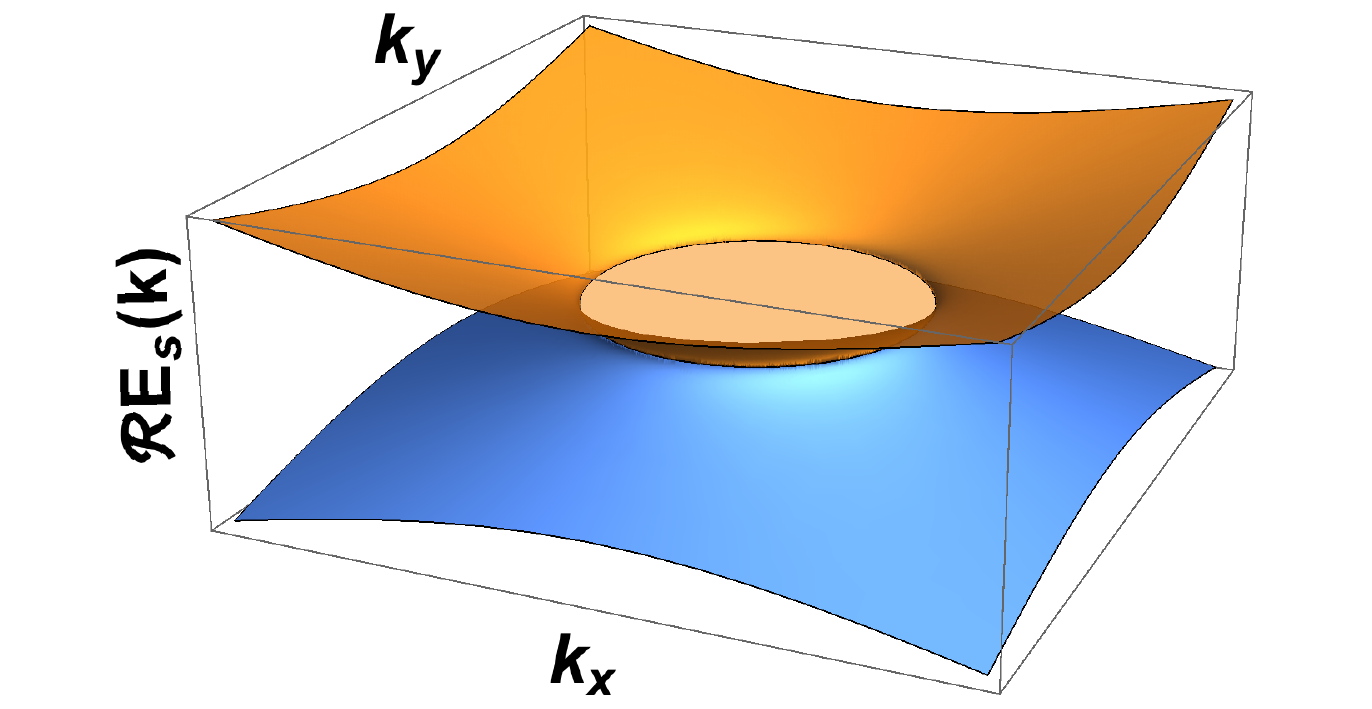}
\end{tabular}
\caption{(Color online) 
The real part of the spectrum $E_s(k_x, k_y, 0)$ of Weyl semimetal in which the tilting of the Weyl cone is along $\boldsymbol{e}_z$ in momentum space. 
(Left) The band touching Weyl point in the absence of disorder. (Right) In the disordered type-II Weyl semimetal the Weyl point expands into a circle flat band, which is defined by $k_x^2+k_y^2\leq 1/4\tau^2 C^2$, where $\tau$ is the mean free time and $C$ is the tilt parameter of the Weyl cone.}
\label{fig2} 
\end{figure}

In this paper, we investigate the interplay of the disorder scattering and the band-structure topology in  
Weyl semimetals with a tilt of a Weyl cone. In particular, we calculate the self-energy correction due to electron elastic scattering on disorder 
to the one-particle Green function within the first Born approximation. We observe that the complex anisotropic dispersion of particles near the type-II 
Weyl point is given by the eigenvalues of the non-Hermitian matrix, which contains contributions from the
Weyl Hamiltonian and complex self-energy. 
We show that for the case of a tilt in a single direction the type-II Weyl point expands into a flat band, lying on the plane perpendicular to the direction of the tilt as shown in Figs. \ref{fig1} and \ref{fig2}. 
However, a more general form of the tilt might shrink the flat band into a nodal line segment. Let us now show that elastic scattering on disorder in type-II Weyl semimetals might affect the topological properties of the electronic band structure.

We consider a minimal model of a time reversal symmetry-breaking Weyl semimetal with only two nodes of 
opposite chirality $s=\pm 1$ in its bandstructure. The linearized Hamiltonian around each Weyl point is given by
\begin{equation}\label{Ham}
H_{s}(\mathbf{k}) = s C k_z+sv\boldsymbol{\sigma}\cdot\mathbf{k}.
\end{equation}
Here $\boldsymbol{\sigma}$ is a vector composed of the three Pauli matrices, $v$ is the Fermi velocity when the tilting parameter is $C=0$, and we set $\hbar=1$ in our calculations. We choose to consider time-reversal symmetry breaking and inversion symmetric tilt such that two Weyl cones are tilted in opposite directions with respect to each other and separated along the $\boldsymbol{e}_z$ axis in momentum space (we will discuss the case of general tilt later in the conclusions). The condition $|C|=v$ defines the point of Lifshitz transition between type-I ($|C|<v$) and type-II ($|C|>v$) phases of the Weyl semimetal.

We consider the short-range impurity scattering potential of the form $V(\mathbf{r})=u_0\sum_a\delta(\mathbf{r}-\mathbf{r}_a)$, where $\mathbf{r}_a$ labels the impurity positions.
We neglect the electron's inter node relaxation rate due to impurity elastic scattering compared to the intra node relaxation rate, assuming that the impurity potential does not mix Weyl fermions at different nodes. We also do not consider Coulomb interaction here. 

To proceed, let us now study the effect of disorder on the behavior of the electron Green function.
The expression for the disorder induced electron self-energy in the first Born approximation is given by
\begin{equation}
\Sigma^{R}_s(\omega) = n_0u_0^2\int \frac{d^3k}{(2\pi)^3}G^{R}_{s}(\omega,\mathbf{k}),
\end{equation}
where $n_0$ is the impurity concentration and the one-particle retarded Green functions have the following form
\begin{equation}
G^{R}_{s}(\omega,\mathbf{k})= \frac{1}{2}\sum_{t=\pm1}\frac{1- st\boldsymbol{\sigma}\cdot\mathbf{n}}{\omega+\mu-sCk_z+t vk+i\delta},
\end{equation}
where $\mathbf{n}=\mathbf{k}/|\mathbf{k}|$ is the unit vector in the direction of momentum $\mathbf{k}$. 
The linearized model is not well applied for computing the self-energy around the Lifshitz
transition between type-I and type-II phases of the Weyl semimetal. Higher-order momentum corrections to the Hamiltonian Eq. \ref{Ham} become dominant, and the linearized model is not valid. Although it might be applicable in two limiting cases of small, $|C|\ll v$, and large, $|C|\gg v$, tilts far from the transition as discussed in Ref. \cite{Yago_Jens}. Hence we proceed by calculating the self-energy due to scattering on disorder in these opposite regimes of the tilt.

In the case $v\gg|C|$ the tilt-dependent correction to the self energy in the linear order in $|C|/v$ has the form
\begin{eqnarray}\nonumber\label{SE_type1}
\Sigma^{R}_s(\omega) = -(\omega+\mu)v\Lambda_d\frac{2\gamma}{\pi}\bigg(1-\frac{2C}{3v}\sigma^z\bigg)\\
-i(\omega+\mu)^2\gamma\bigg(1-\frac{C}{v}\sigma^z\bigg),
\end{eqnarray}
where $\gamma=\frac{n_0u_0^2}{4\pi v^3}$. The tilt modifies the matrix structure of the self-energy. The real part of the self-energy is proportional to $\omega+\mu$, which expresses the renormalization of the spectral weight. The divergence of the real part is an artifact of the model of the $\delta$-function impurity potential.
Hence, we introduce a momentum cut-off $\Lambda_{d}$, which is proportional to the characteristic inverse half-width of the more realistic finite range scattering potential and assume that $v\Lambda_d\gg |\omega+\mu|$. The imaginary part defines the electron's inverse mean free time due to elastic scattering on disorder. Finally, the assumption that the impurity potential does not scatter particles between two Weyl nodes, is fulfilled provided that the distance between two Weyl points in momentum space is much larger than $\Lambda_d^{-1}$.

For the case of a large tilt deep in the type-II Weyl semimetal phase $|C|\gg v$ the self-energy in the linear order in $v/|C|$ is given by
\begin{eqnarray}\label{SE_type2}\nonumber
\Sigma^{R}_s(\omega) &=&\frac{2}{\pi\tau}\frac{\omega+\mu}{|C|\Lambda}\bigg(1-\frac{3v}{2C}\sigma^z\bigg) -\frac{i}{2\tau}\bigg(1-\frac{v}{C}\sigma^z\bigg)\\
&+&\frac{i}{2\tau}\bigg[\frac{\omega+\mu}{|C|\Lambda}\bigg]^2\bigg(1+\frac{3v}{C}\sigma^z\bigg),
\end{eqnarray}
where $\tau^{-1} = n_0u_0^2\Lambda^2/4\pi |C|$.
Here both the real and the imaginary parts of the self-energy have divergency with the momentum cut off $\Lambda\gg \Lambda_d$. 
This is now a signature of the presence of unbounded electron-hole pockets in the linearized model. In the computation of the expression in Eq. \ref{SE_type2} 
we consider the limit $|C|\Lambda \gg |\omega+\mu|$, in which the density of states at the Fermi level is determined by the width of the pockets. 
Hence the real part and the imaginary part on the second line in Eq. \ref{SE_type2} are small correction to the self-energy and will be neglected in what follows.  

We are now in the position to find the energy and the decay rate of the electrons by calculating the poles of the disorder-averaged retarded Green function,
\begin{equation}
\mathrm{det}[\{{G^{R}_s}(\omega,\mathbf{k})\}^{-1}-\Sigma^{R}_s(\omega) ]=0,
\end{equation}
where $\mathrm{det}$ denotes the determinant of the matrix. In the case $v\gg |C|$ we obtain an equation for the dispersion of particles $E_{s}(\mathbf{k})$ in the form
\begin{eqnarray}\label{Spectrum_type1}
&\bigg\{&sCk_z -[1+\frac{2\gamma v}{\pi}\Lambda]E_{s}(\mathbf{k})-i\gamma E_{s}^2(\mathbf{k})\bigg\}^2\\\nonumber
&=&v^2k_{\perp}^2 + \bigg[svk_z-\frac{4\gamma}{3\pi}C\Lambda E_{s}(\mathbf{k})-i\frac{C}{v}\gamma E_{s}^2(\mathbf{k})\bigg]^2,
\end{eqnarray}
where $k_{\perp}=\sqrt{k_x^2+k_y^2}$.  
The physical solution of the forth degree equation Eq. \ref{Spectrum_type1} gives the spectrum of particles in the disordered type-I Weyl semimetal
$E^{(\pm)}_{s}(\mathbf{k}) \approx \pm (sCk_z\pm vk)/[1+\frac{2\gamma v}{\pi}\Lambda]-i\gamma v^2k^2_{\perp}/[1+\frac{2\gamma v}{\pi}\Lambda]^3$.

\begin{figure}[t]
\includegraphics[width=8.0cm]{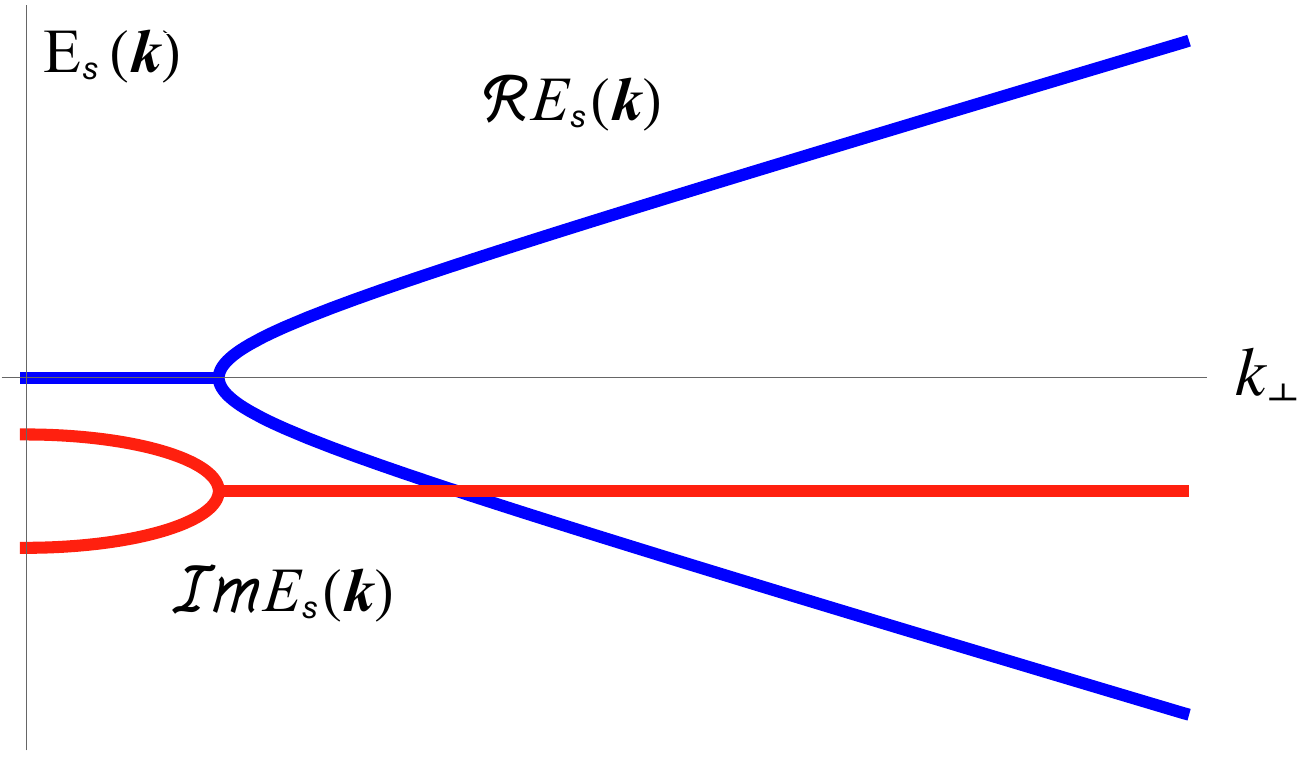}
\caption{(Color online) 
 The real and imaginary parts of the spectrum $E_s(\mathbf{k})$ at $k_z=0$ as a function of momentum $k_{\perp}=\sqrt{k_x^2+k_y^2}$ of disordered type-II Weyl semimetal. The spectrum is purely imaginary within the region of momenta 
 $0\leq k_{\perp}\leq 1/2|C|\tau$, where $\tau$ is the mean free time and $C$ is the tilt parameter. At $k_{\perp}= 1/2|C|\tau$ the spectrum has a single imaginary value of $-i/2\tau$.}
\label{fig3}
\end{figure}

Let us now see what happens with the dispersion of particles in the disordered type-II Weyl semimetal in the limit $|C|\gg v$. We find that the spectrum acquires an unusual complex form
\begin{eqnarray}\label{Main_Result}
E_{s}^{(\pm)}(\mathbf{k})=sCk_z
\pm v\sqrt{k_{\perp}^2+\bigg(sk_z+\frac{i}{2C\tau}\bigg)^2}-\frac{i}{2\tau}.
\end{eqnarray}
At momentum $k_z=0$, expression for $E_{s}^{(\pm)}(\mathbf{k})$ in Eq. \ref{Main_Result} reduces to
\begin{equation}
E_{s}^{(\pm)}(\mathbf{k})|_{k_z=0}=\pm v\sqrt{k_{\perp}^2-\frac{1}{4C^2\tau^2}} -\frac{i}{2\tau}.
\end{equation}
We find that $E_{s}^{(\pm)}(\mathbf{k})|_{k_z=0}$ is fully imaginary and dispersionless in the region of momenta $0\leq k_{\perp}\leq 1/2|C|\tau$, see Figs. \ref{fig2} and \ref{fig3}
(we note here that the distance between two Weyl points of opposite chirality in momentum space is considered to be much larger than $1/|C|\tau$, such that two flat bands do not overlap).
Generally, the imaginary part of the spectrum $\Im E_{s}^{(\pm)}(\mathbf{k})|_{k_z=0}$ has two values in the region $0\leq k_{\perp}< 1/2|C|\tau$ and a single value in $k_{\perp}\geq 1/2|C|\tau$. 
The spectrum contains an \emph{exceptional nodal ring} at momenta $\mathbf{k}_e$, which are defined by the equations $k_x^2+k_y^2=1/4C^2\tau^2$ and $k_z=0$. An exceptional nodal ring separates the flat band and
the dispersive parts of the spectrum. The dispersion relation at the exceptional nodal ring has a single value of $-i/2\tau$, which is purely imaginary. Hence the matrix of the disorder averaged Green function in the momentum-frequency representation at $\mathbf{k}=\mathbf{k}_e$ is defective,
\begin{equation}
\langle G_s^{R}(\omega,\mathbf{k}_e)\rangle = \frac{\omega+sv\boldsymbol{\sigma}\cdot\mathbf{k}_e+\frac{i}{2\tau}[1+\frac{v\sigma^z}{C}]}{(\omega+\frac{i}{2\tau})^2}.
\end{equation}
There is a square-root singularity in the vicinity of the exceptional nodal ring $E_{s}^{(\pm)}(\mathbf{k}_e+\mathbf{k})\approx\pm v\sqrt{|C|k_{\perp}-1/2\tau+2isCk_z}/|C|\sqrt{2\tau} -\frac{i}{2\tau}$, which bears a close resemblance to the behavior of the spectrum around the exceptional points in two-dimensional optic and electronic systems \cite{NonHerm1,NonHerm4,flach,NonHerm2,NonHerm3,Shen_Fu, Kozii_Fu}. Note however that the spectra of the flat bands on the surfaces of the three-dimensional Dirac systems \cite{Heikila_Volovik, BHB} have no such singularity.

We also compare our results with the case of the Weyl system formed out of cold atom gases with the gain and loss for spin-up and spin-down atoms  \cite{Weyl_ring}. 
In a such system, the imaginary part of the spectrum at the flat band has positive and negative values on the imaginary axis and vanishes at the exceptional nodal ring. In our case the spectrum is contained on the lower half-plane of the complex plane, and the flat band is generated in equilibrium due to the interplay of the disorder scattering and the tilt of the Weyl cone. The physical origin of the flat band in the type-II Weyl semimetal case is the suppression of the probability of scattering on the impurity within the plane perpendicular to the direction of the tilt.

It is important to stress that our calculations are performed for the case when the Weyl cone is tilted along one direction in momentum space. The general case of the tilt can be described by
the linearized Hamiltonian around each Weyl point $H_{s}(\mathbf{k}) = s [\mathbf{C} +v\boldsymbol{\sigma}]\cdot\mathbf{k}$, where $\mathbf{C}$ is a vector. For example the spectrum of particles in a disordered type-II Weyl semimetal for $|C_{x,y}|\equiv|C|>v$ and $C_z=0$ is given by $ E_{s}^{(\pm)}(\mathbf{k})=s\mathbf{C}\cdot\mathbf{k} 
\pm v\{k_z^2+\sum_{n=x,y}(sk_n+ i/2C\tilde{\tau})^2\}^{1/2}- i/2\tilde{\tau}$. Hence depending on the values of the tilt parameters either a flat band or a line segment of nodes might be realized in the bandstructure. The line segment with exceptional nodal edge points is a three-dimensional analog of the Fermi arc studied in Ref. \cite{Kozii_Fu}.

Let us now discuss the effect of tilt on the dispersion of surface states.
We consider a boundary at $x=0$ of a clean Weyl semimetal. In the volume of the Weyl semimetal $x>0$, 
using the real-space representation of the Hamiltonian $H_{s}(\mathbf{k}) = s [\mathbf{C} +v\boldsymbol{\sigma}]\cdot\mathbf{k}$, we search for the wave-function of the surface state in the form $\Psi_s(\mathbf{r}) =\exp[-\alpha_s x+i(k_{y}y+k_{z}z)](A_s,B_s)^{\mathrm{T}}$, where $\Re\alpha_s >0$ and $A_s, B_s$ are some constants. 
We apply a condition of the zero current through the surface $x=0$, which gives 
\begin{equation}\label{zero_current}
\int dS \Psi_s^{+}(\mathbf{r})(C_x+v\sigma_x)\Psi_s(\mathbf{r})|_{x=0}=0,
\end{equation}
where integration is performed over the surface of the system.
The profound discussion of the problem of boundary conditions on the surface states in Dirac materials is given in \cite{Enaldiev}.
The general boundary condition for the two-component wave function has the form $(1,-\gamma_s)\Psi_s(\mathbf{r})|_{x=0}=0$, where the complex parameter $\gamma_s(y,z)\equiv a_s + i b_s$ in which $a_s, b_s\in \Re$, might depend on the position at the boundary (here we consider $\gamma_s$ to be constant).  This constraint determines the wave function at the boundary 
$
\Psi_s(\mathbf{r})|_{x=0}=\mathcal{N}(\gamma_s,1)^{\mathrm{T}}
$, which must nullify Eq. \ref{zero_current}  (here $\mathcal{N}$ is the normalization factor). Therefore we obtain the equation, 
\begin{equation}\label{BC}
(a_sC_x+ v)^2+b_s^2C_x^2=v^2-C_x^2
\end{equation}
to calculate parameters $a_s$ and $b_s$. We see that the boundary condition can not be satisfied for $|C_x|>v$ and hence there are no surface states. This result is compatible with Ref. \cite{Goerbig}. In the 
opposite case of $v>|C_x|$, the solution for Eq. \ref{BC} can be parametrized by angle $\theta_s$ as
$b_s= \sqrt{v^2/C_x^2-1}\sin\theta_s$ and $a_s= -v/C_x +\sqrt{v^2/C_x^2-1}\cos\theta_s$, such that we can rewrite 
\begin{equation}
\gamma_s=-\frac{v}{C_x} +\left[\frac{v^2}{C_x^2}-1\right]^{1/2}\exp(i\theta_s).
\end{equation}
Using the constraint for the wave function at the boundary, we find the equation for the complex parameter $\alpha_s =\frac{1}{1-\gamma_s^2}[2i\gamma_s k_z+(1+\gamma_s^2)k_y]$. The condition $\Re \alpha_s>0$ determines the region of momenta in which the boundary solution exists. 
In the limit of small tilt $|C_x|/v\ll 1$ we observe that $|b_s|\approx 1$, $|a_s|\ll 1$ and find the inverse localization length $\Re\alpha_s = [(1-b_s^2)k_y-2b_sk_z]/(1+b_s^2)$, the parameter which describes the oscillations of the surface-state wave-function $\Im\alpha_s = [2b_sk_y+(1-b_s^2)k_z]\frac{2a_s}{(1+b_s^2)^2}$ and the dispersion,
\begin{eqnarray}
E^{\mathrm{surf}}_s(\mathbf{k}) = sC_yk_y+sC_zk_z-vs\frac{1+b_s^2}{2a_s}\Im\alpha_s
\end{eqnarray}
In the limit of large tilt $|C_x|/v\rightarrow 1$ at the vicinity of the transition we find $\alpha_s = [k_y\mathrm{sgn}(C_x)-ik_z]\frac{\exp[-i\theta_s]}{\sqrt{2(1-|C_x|/v)}}$ and the spectrum $E^{\mathrm{surf}}_s = sC_yk_y+sC_zk_z-s\sqrt{v-|C_x|}[k_y\sin\theta_s+k_z\cos\theta_s \mathrm{sgn}(C_x)]$. 
To summarize, we find that the localization length becomes zero and the spectral region vanishes in the limit $|C_x|\rightarrow v$, so that the surface states do not exist in the case $|C_x|>v$. The above discussion qualitatively holds for the disordered Weyl semimetal as long as the localization length of the surface states is much smaller than the electron mean free path in the bulk.

It is also instructive to comment on the possible physical realization of the tilted conical spectrum around the Weyl point. An accurate description of the Weyl cone tilting effects might be considered within the model of a Weyl semimetal based on the topological insulator - ferromagnetic insulator multilayer heterostructure \cite{Burkov_Balents}.
There the separation of two Weyl points is accomplished by breaking time-reversal symmetry due to the exchange 
coupling of the electron spin and localized spins in the ferromagnet. We argue that taking into account inevitable spin-dependent inter layer tunneling effects in addition to spin-independent tunneling leads to the tilt of the spectrum around the Weyl cones. However, this model might not be general since the tilting of the spectrum can be engineered along the direction of separation of the Weyl points. 
The details of this proposal and the study of the nonlinear band structure effects around the transition between type-I and type-II Weyl semimetals on moving and merging of the flat bands is beyond the scope of the present paper.

To conclude, we studied the change in the topology of the electronic band structure of a Weyl semimetal by disorder. In particular, we found that in disordered type-II Weyl semimetal the Weyl point expands into a flat band. This results from the anisotropic spectrum of type-II Weyl fermions and the matrix structure of the complex self-energy due to electron scattering on disorder. 
The interplay of disorder and topological properties of the electronic bandstructure might be studied in $\mathrm{MoTe_2}$ and $\mathrm{TaIr Te_4}$, which show experimental signatures of the type-II Weyl semimetal phase (for a review, see Ref. \cite{Weyl_review}). 

Finally, it would be interesting to extend the above-presented research to the semimetals with tilted nodal line spectra \cite{BHB}. 
In particular, we expect the nodal line to acquire a finite width due to disorder, i.e. the nodal line transforms into a flat band of finite width bounded by the exceptional nodal lines.

\begin{acknowledgements}
We thank V. Kozii, T. Ojanen, and V. Zyuzin for stimulating discussions and Egor Babaev for his warm hospitality at the Royal Institute of Technology. We also thank J. Nissinen and G. Volovik for bringing Ref. \cite{Weyl_ring} to our attention. A.A.Z. was supported by the Academy of Finland (project No. 308339) and, in part, by the Swedish Research Council Grant No. 642-2013-7837.
\end{acknowledgements}

\bibliography{Disorder_Weyl4}
\end{document}